\documentclass[acmsmall,screen]{acmart}
\AtBeginDocument{%
  }

\acmConference[Conference acronym 'XX]{Make sure to enter the correct
  conference title from your rights confirmation emai}{June 03--05,
  2018}{Woodstock, NY}

\usepackage{tablefootnote}
\usepackage{tikz}
\usepackage{amsmath}
\usepackage{pifont}
\usepackage{stackengine}
\usepackage{filecontents}
\usepackage{colortbl}%
\newcommand{\grayrow}{\rowcolor[gray]{0.925}}
\usepackage{enumitem}
\setlist{itemsep=1pt,topsep=1pt,parsep=0pt,partopsep=0pt}
\usepackage{fontawesome}

\newcommand*{\tool}{{\textsc{Verite}}\xspace}
\newcommand*{\toolnoaction}{{\textsc{Verite-NOACT}}\xspace}
\newcommand*{\toolnoabnormal}{{\textsc{Verite-NOCDT}}\xspace}
\newcommand*{\toolnoaccount}{{\textsc{Verite-NOACC}}\xspace}
\newcommand*{\toolnograd}{{\textsc{Verite-NOGRD}}\xspace}

\newcommand{\tab}{\hspace*{1em}}

\newcommand{\ityfuzz}{\textsc{ItyFuzz} }

\usepackage{url}
\usepackage{ wasysym }
\usepackage{ makecell }
\usepackage{tabularx}
\usepackage{multirow}
\usepackage{listings}
\usepackage{xspace}
\usepackage{longtable}
\usepackage{xtab}
\usepackage{booktabs}
\usepackage{subcaption}
\usepackage[utf8]{inputenc}
\usepackage[T1]{fontenc}
\usepackage{mathtools}
\usepackage[thinc]{esdiff}
\usepackage{algpseudocode}
\usepackage{array,amsmath}
\usepackage{bm}
\usepackage{subcaption}
\usepackage[ruled,vlined,linesnumbered]{algorithm2e}

\SetCommentSty{mycommfont}
\definecolor{codegreen}{rgb}{0,0.6,0}
\definecolor{codegray}{rgb}{0.8,0.5,0.5}
\definecolor{codepurple}{rgb}{0.58,0,0.82}
\definecolor{backcolour}{rgb}{0.95,0.95,0.92}

\newcommand{\compactline}{\looseness=-1}

\newcommand{\vulna}{\textsc{ENDERV1}\xspace}
\newcommand{\vulnb}{\textsc{ETHOSX}\xspace}

\lstdefinestyle{mystyle}{
    backgroundcolor=\color{backcolour},   
    commentstyle=\color{codegreen},
    keywordstyle=\color{magenta},
    numberstyle=\tiny\color{codegray},
    stringstyle=\color{codepurple},
    basicstyle=\ttfamily\footnotesize,
    breakatwhitespace=false,         
    breaklines=true,                 
    captionpos=b,                    
    keepspaces=true,                 
    numbers=left,                    
    numbersep=5pt,                  
    showspaces=false,                
    showstringspaces=false,
    showtabs=false,                  
    tabsize=2,
    xleftmargin=12pt,
    otherkeywords={function, external, internal, address, uint256}
}

\setcopyright{cc}
\setcctype{by-nc-nd}
\acmDOI{10.1145/3715720}
\acmYear{2025}
\acmJournal{PACMSE}
\acmVolume{2}
\acmNumber{FSE}
\acmArticle{FSE008}
\acmMonth{7}
\received{2024-09-13}
\received[accepted]{2025-01-14}

\begin{document}

\title{Smart Contract Fuzzing Towards Profitable Vulnerabilities}

\author{Ziqiao Kong}
\orcid{0009-0009-4926-4932}
\affiliation{%
  \institution{Nanyang Technological University}
  \city{Singapore}
  \country{Singapore}
}
\email{ziqiao001@e.ntu.edu.sg}

\author{Cen Zhang}
\orcid{0000-0001-5603-1322}
\affiliation{%
  \institution{Nanyang Technological University}
  \city{Singapore}
  \country{Singapore}
}
\email{cen001@e.ntu.edu.sg}
\thanks{Cen Zhang is the corresponding author.}

\author{Maoyi Xie}
\orcid{0009-0001-4496-5037}
\affiliation{%
  \institution{Nanyang Technological University}
  \city{Singapore}
  \country{Singapore}
}
\email{maoyi001@e.ntu.edu.sg}

\author{Ming Hu}
\orcid{0000-0002-5058-4660}
\affiliation{%
  \institution{Singapore Management University}
  \city{Singapore}
  \country{Singapore}
}
\email{minghu@smu.edu.sg}

\author{Yue Xue}
\orcid{0009-0004-2141-2044}
\affiliation{%
  \institution{MetaTrust Labs}
  \city{Singapore}
  \country{Singapore}
}
\email{xueyue@metatrust.io}

\author{Ye Liu}
\orcid{0000-0001-6709-3721}
\affiliation{%
  \institution{Singapore Management University}
  \city{Singapore}
  \country{Singapore}
}
\email{yeliu@smu.edu.sg}

\author{Haijun Wang}
\orcid{0009-0001-3509-3919}
\affiliation{%
  \institution{Xi'an Jiaotong University}
  \city{Xi'an}
  \country{China}
}
\email{haijunwang@xjtu.edu.cn}

\author{Yang Liu}
\orcid{0000-0001-7300-9215}
\affiliation{%
  \institution{Nanyang Technological University}
  \city{Singapore}
  \country{Singapore}
}
\email{yangliu@ntu.edu.sg}

\renewcommand{\shortauthors}{Ziqiao Kong et al.}


\begin{abstract}

Billions of dollars are transacted through smart contracts, making vulnerabilities a major financial risk.
One focus in the security arms race is on profitable vulnerabilities that attackers can exploit.
Fuzzing is a key method for identifying these vulnerabilities.
However, current solutions face two main limitations: 
\ding{172} a lack of profit-centric techniques for expediting detection and 
\ding{173} insufficient automation in maximizing the profitability of discovered vulnerabilities, leaving the analysis to human experts.

To address these gaps, we have developed \tool{}, a profit-centric smart contract fuzzing framework that not only effectively detects those profitable vulnerabilities but also maximizes the exploited profits.
\tool{} has three key features:
\ding{172} DeFi action-based mutators for boosting the exploration of transactions with different fund flows;
\ding{173} potentially profitable candidates identification criteria, which checks whether the input has caused abnormal fund flow properties during testing;
\ding{174} a gradient descent-based profit maximization strategy for these identified candidates.
\tool{} is fully developed from scratch and evaluated on a dataset consisting of 61 exploited real-world DeFi projects with an average of over 1.1 million dollars loss.
The results show that \tool can automatically extract more than 18 million dollars in total and is significantly better than state-of-the-art fuzzer \ityfuzz in both detection (29/10) and exploitation (134 times more profits gained on average).
Remarkably, in 12 targets, it gains more profits than real-world attacking exploits (1.01 to 11.45 times more).
\tool{} is also applied by auditors in contract auditing, where 6 (5 high severity) zero-day vulnerabilities are found with over \$2,500 bounty rewards.

\end{abstract}



\maketitle

\section{Introduction}

%
%


Smart contracts are infrastructures for Web 3.0 businesses, managing billions of dollars in digital assets.
However, they are also susceptible to vulnerabilities such as logic bugs, which can potentially lead to severe financial losses~\cite{zhou2023sok}.
A notable fact about these vulnerabilities is that they are often economically profitable, making them highly attractive to malicious actors.
Regrettably, exploits of these susceptibilities are becoming a pervasive issue, with successful attacks occurring on even a daily basis, where hackers illicitly profit from these weaknesses.
For example, in the past first month of 2024, there have been nine DeFi (Decentralized Finance) hack incidents with a total loss of over 100 million dollars~\cite{defihacklabs}.
Generally, detecting, exploiting, and mitigating the smart contract vulnerabilities have been an arm race between malicious actors and whitehats, especially for the detection of those vulnerabilities that attackers could exploit to gain profits.
\compactline

Fuzzing is an effective vulnerability detection approach.
Basically, smart contract fuzzing is a process that generates inputs to execute the APIs of a target contract and explores transaction sequences to detect abnormal behaviors that indicate bugs.
Mainstream fuzzers, such as those mentioned in \cite{ityfuzz, efcf, echidna}, operate at the API level, combining transaction sequences for comprehensive testing.
Different fuzzers adopt varied techniques: sFuzz \cite{sfuzz} focuses on higher code coverage, Smartian \cite{smartian} uses dataflow analysis, and ConFuzzius \cite{confuzzius} integrates symbolic execution and taint analysis.
Additionally, \ityfuzz \cite{ityfuzz} uses dataflow waypoints to speed up fuzzing, while learning-based methods are employed by tools like ILF \cite{he2019learning}, xFuzz \cite{xue2022xfuzz}, and RLF \cite{rlf}, showcasing the diverse strategies in this field.
\compactline

Standing from the perspective of fuzzing towards profitable vulnerabilities, existing fuzzers have shown certain limitations.
\textbf{First, there is a lack of efforts on boosting the fuzzing towards these profitable vulnerabilities.}
Currently, mainstream fuzzers are code-centric, with strategies primarily focused on maximizing code coverage.
While this approach is generally effective, it does not capitalize on the specific characteristics of profitable vulnerabilities.
Such vulnerabilities often contain issues like price manipulation, inadequate access control, or other misbehaviors closely tied to contract-specific business logic, which are not always detectable via its code features like coverage.
From another perspective, the transaction fund flow features can provide complementary or unique insights, enabling fuzzers to more effectively target these vulnerabilities.
For instance, a profitable vulnerability implies that there is a transaction where the fund flow demonstrates that an attacker ultimately profits.
Current fuzzers do not effectively leverage these characteristics to expedite the detection of profitable vulnerabilities.
\textbf{Second, existing fuzzers halt after triggering the vulnerabilities, leaving further exploitation to human analysts.}
Most fuzzers are content with producing a proof-of-concept (PoC) that triggers a vulnerability without pursuing further utilization.
In reality, the automatic exploitation of profitable vulnerabilities is crucial for the security of smart contracts.
If security experts or auditors could automatically generate inputs that maximize profit, it would significantly enhance their understanding of the severity and ranking of the vulnerability and subsequent mitigation strategies.
Unfortunately, even when current fuzzers trigger a vulnerability and produce a PoC, they do not proceed to maximize profit potential, leaving a gap between the vulnerability and exploitation\cite{vulnerablenotexploirted}.
This gap indicates a missed opportunity to fully leverage fuzzers to enhance smart contract security.
\compactline

To address these limitations, we propose a profit-centric fuzzing framework, \tool{}, which not only targets the discovery of profitable vulnerabilities but also maximizes the exploitation of profits.
Our framework incorporates three key designs:
\ding{172} \textbf{Action-Based Mutators}: Instead of relying solely on API-level mutations, we introduce action-based mutators to enhance the exploration of fund flow spaces.
DeFi actions, which are semantic groups of contract APIs, represent logical transaction steps, such as token swaps or transfers.
By mutating at the action level, \tool{} generates more valid transactions compared to API-level mutations, which often disrupt transaction semantics.
We have summarized 10 common actions from mainstream DeFi projects and developed action-based mutators to accelerate this exploration.
\ding{173} \textbf{Profitability Recognition}: \tool{} identifies potentially profitable candidates by examining whether inputs trigger abnormal fund flow properties during testing.
It evaluates four key properties: net positive profit across all assets, imbalance in Uniswap pairs, unconditional token gains, and unconditional token burns.
If any property is satisfied, \tool{} considers it a candidate and begins the profit maximization process.
\ding{174} \textbf{Profit Maximization}: We model profit maximization as a gradient descent problem and devise a corresponding strategy for optimizing identified candidates.
The intuition is that these candidates already have the basis to make profits, \textit{i.e.,} potentially profitable transaction sequences.
The next step is to refine the API parameters within these sequences to maximize profits.
\tool{} models the profit functions as $f(x_1, x_2, \ldots)$, with API parameters as arguments $x$, and employs the gradient descent algorithm to find the optimal parameters.
\compactline

To thoroughly evaluate \tool{}, we expanded the existing \ityfuzz dataset to include 61 open-source Web3 projects by collecting the latest DeFi hack incidents.
Each project features a known attack exploiting a profitable vulnerability, with an average loss of approximately \$1.1 million.
In our evaluation, \tool{} demonstrated superior ability in detecting profitable vulnerabilities and achieved significantly higher profits in generated exploits compared to the state-of-the-art fuzzer \ityfuzz.
Specifically, \tool{} successfully exploited 29 targets by yielding 18 million profits, whereas \ityfuzz managed only nine with 100 thousand profits.
On targets found by both fuzzers, \tool{} discovered, on average, 58 times more profits than \ityfuzz.
Remarkably, when compared to real-world attack exploits, typically optimized by attackers, \tool{} achieved higher profits on 12 targets, with profit ratios ranging from 1.01 to 11.45 times.
Our in-depth analysis further elucidates these comparisons, and our ablation study confirms that each key design component of \tool{} significantly enhances its overall performance.
To demonstrate its practicality, \tool{} was employed to assist auditors in detecting and exploiting zero-day vulnerabilities on popular Web3 auditing contest platforms Secure3~\cite{secure3}.
\tool{} uncovered six vulnerabilities, five of which were high severity.
We responsibly reported these issues, assisted in their fix, and received over \$2,500 in bounty rewards.
\compactline

In summary, our contributions are:
\begin{itemize}
\item We first identified the gaps and challenges for building fuzzers towards profitable vulnerabilities, including the lack of profit-centric design and the absence of profit maximization.
\compactline
\item We proposed three key designs for addressing these gaps, including action-based mutators, fund-flow-property-based profitable candidate recognition, and gradient descent-based profit maximization.
We further implemented \tool{} from scratch.
\compactline
\item We built a comprehensive evaluation dataset and conducted extensive experiments to evaluate \tool{}'s performance. \tool{} shows superior performance compared to existing solutions and can even gain more profits than real-world attacking exploits (11.45 times more in the best case).
\compactline
\item \tool{} has been applied to help auditors find six zero-day profitable vulnerabilities with \$2,500+ bug bounty, showing its practical value.
\compactline
\end{itemize}


\section{Preliminaries}

\subsection{Concept Modeling}

\noindent
\textbf{Smart Contracts and Transactions} \tab Smart contracts are usually written in a programming language like solidity, compiled into EVM byte code, and deployed on EVM~\cite{evm}.
A transaction interacting with smart contracts carries several important fields, including \verb|to| specifying the target contract address, \verb|value| specifying the ether transferred, and \verb|data| encoding which smart contract function the sender wishes to call and corresponding arguments.
Therefore, a single transaction corresponds to one smart contract API function call, and a transaction sequence consists of several transactions in a fixed order.
In addition, each contract function can also contain any number of transactions, sometimes also called "internal transactions", to any other smart contracts due to its turning completeness. 
\compactline

The execution of a transaction on EVM can have two results: \textit{return} or \textit{revert}. If a transaction stops and \textit{returns} normally, it will then commit the new state to the EVM. Otherwise, the current state will be discarded, and EVM will \textit{revert} to the previous state. 
\compactline

\noindent
\textbf{Attack Contract} \tab The attack contract is usually a contract that an attacker deploys as an agent to send transactions to target contracts to exploit vulnerabilities. The benefit of deploying an attack contract and calling the attack contract compared to sending several transactions is that the attacker can revert the transaction anytime to avoid any loss and maximize profits. This also aligns with the observation of the trending of flash loan usages on the blockchain~\cite{flashloan}, which requires users to provide a deployed contract.
\compactline


\begin{lstlisting}[float=t, language=Python, caption={The simplified implementation of swapping two tokens using a UniswapV2 router (Action A2 in Table \ref{tab:actions}).}, label={lst:action}]
function swap2_action(address token_from, address token_to, uint percentage) {
    uint amount = token_from.balanceOf(address(this)) * percentage / 100;

    token_from.approve(Router, amount); 
    address[] memory path = new address[](2);
    path[0] = token_from;
    path[1] = token_to;
    Router.swapExactETHForTokensSupportingFeeOnTransferTokens(amount, path, ...);
}
\end{lstlisting}

\begin{figure}[t]
    \centering
    \includegraphics[width=\linewidth]{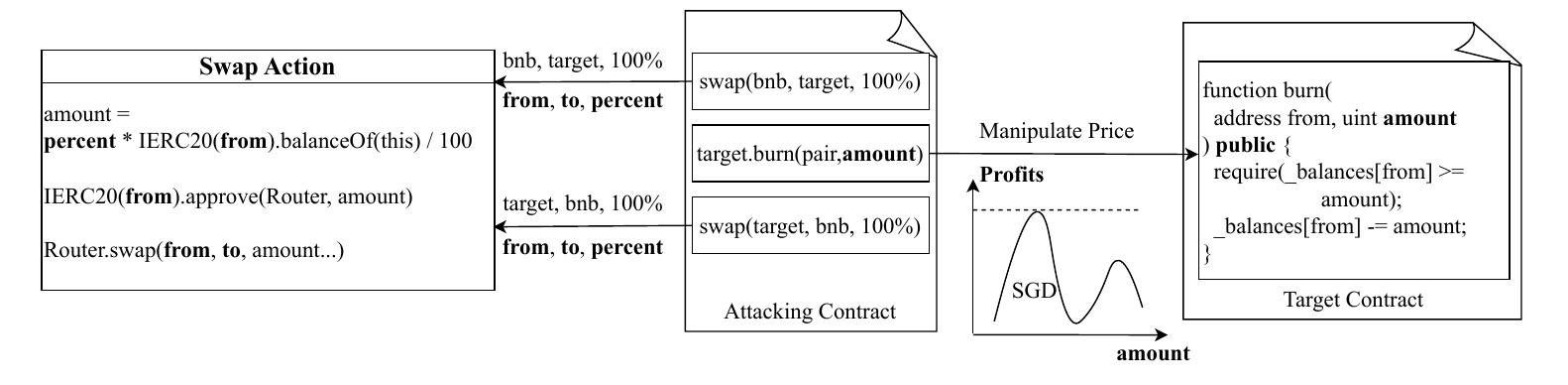}
    \caption{The running example simplified from one of the vulnerabilities in our real-world application.}
    \label{fig:running}
\end{figure}

\noindent
\textbf{Action} \tab
DeFi is usually implemented with several smart contracts and provides API for external usages. However, there is a gap between the smart contracts functions and the actual high-level business models~\cite{defiranger}: standing from the perspective of a DeFi user, the intention of using a smart contract is usually to complete specific actions, for example, swapping tokens, minting NFTs, etc. We refer to a sequence of smart contract API calls and control flows as "actions", representing users' interactions with DeFi contracts. The Listing~\ref{lst:action} shows a simplified action implementation of swapping two tokens using Uniswap V2. Instead of calling \verb|Router| directly, this action ensures the number of tokens to swap is valid by computing the percentage of the current balance and approves \verb|Router| to do the swap.
The benefit of actions is to extract the high-level DeFi semantics from the low-level smart contracts API and action-based mutation can help the fuzzer generate input at a higher level of semantic granularity comparing with the API-based mutation.
More details about utilization are detailed in Section~\ref{sec:method}.
\compactline

\noindent
\textbf{Gradient Descent} \tab
Gradient descent is widely used in the machine learning area to compute the minimum value of the error function~\cite{gradients}.
Generally, the core idea is that if a function $f(\bm{x})$ is differentiable at a point $\bm{p}$, by continuously moving the point to $\bm{p}^{\prime} = \bm{p} - \alpha \frac{\partial f(\bm{x})}{\partial \bm{x}}$, the value of the function $f(\bm{x})$ shall finally coverage to a minimal value where the gradients $\frac{\partial f(\bm{x})}{\partial \bm{x}}$ is zero.
In practice, a more commonly used variant is Stochastic Gradient Descent (SGD), which chooses the gradient of a random component $x_{i}$ from $\bm{x}$ instead of the full gradients.









\subsection{Running Example}\label{sec:running}



To illustrate the basic idea of how \tool{} finds profitable vulnerabilities and exploits them, we provide an adapted example from a real-world project \vulna where \tool can maximize profits to highlight its key designs in Figure \ref{fig:running}.
The target contract on the right is vulnerable because the \verb|burn| function is wrongly declared as public, allowing anyone to burn other users' tokens, which typically can lead to a price manipulation attack.
In the following, we will explain how the attack can be generated by \tool and why existing solutions are limited.

\noindent
\textbf{D1: Action-Based Transaction Sequence Generation for Effective Mutation} \tab
To exploit this price manipulation vulnerability, an attacker needs to accomplish three strategic steps:
first, acquiring the target tokens through a swap as shown in the attack contract;
second, using the \verb|burn| function to manipulate the market;
Finally, execute a second swap to reverse the initial trade and realize profits, as depicted in the middle of the figure.
Although these steps are essential for exploiting the vulnerability, current fuzzing strategies struggle to generate such transaction sequences because their mutators fail to handle low-level API control and data flow dependencies, resulting in invalid transactions that are often reverted.
For instance, as shown in the left side of the figure, correctly constructing one swap action involves initializing and setting parameters for \verb|Router.swap|, such as ensuring the \verb|amount| is within users' balance and calling \verb|ERC20.approve|.
Existing fuzzing mutators do not possess this detailed understanding, leading to random combinations that fail to create effective transactions, let alone the strategic three-step sequence needed for this exploit.
Based on these observations, we abstracted the common action logic necessary for exploiting profitable vulnerabilities and designed action-level mutators for the fuzzer.
This approach significantly enhances the effectiveness of fuzzing by enabling more efficient exploration of different fund flows during the testing process.
\compactline

\noindent
\textbf{D2: Transaction Parameter Optimization for Profit Maximization} \tab
Current fuzzers often stop exploiting vulnerabilities once they confirm a positive net profit for the attacker.
This approach limits automation in vulnerability assessment and leaves more work for security experts.
By aiming to maximize an attacker's profit during testing, we can automate vulnerability severity evaluation, speed up response times in attack-defense scenarios, and streamline mitigation processes.
This is a valuable area that existing efforts have overlooked.
However, this approach brings new challenges.
For example, the relationship between profit and parameter values is usually non-linear.
To maximize profit, we must consider real-time blockchain states, current token statuses, and the constraints within the target contract's code.
We model this problem as an optimization task to find the maximum value of a continuous function.
For potentially profitable transaction sequences, we treat the fuzzer's controllable variables as function parameters.
Using gradient descent, we aim to find the maximum value, and discover the most profitable exploit.
This method not only improves fuzzing effectiveness but also offers deeper insights into vulnerabilities, enabling faster and more automated defense strategies.
\compactline

\begin{figure*}[t]
    \centering
    \includegraphics[width=1.0\linewidth,page=1, trim=0 0 0 0]{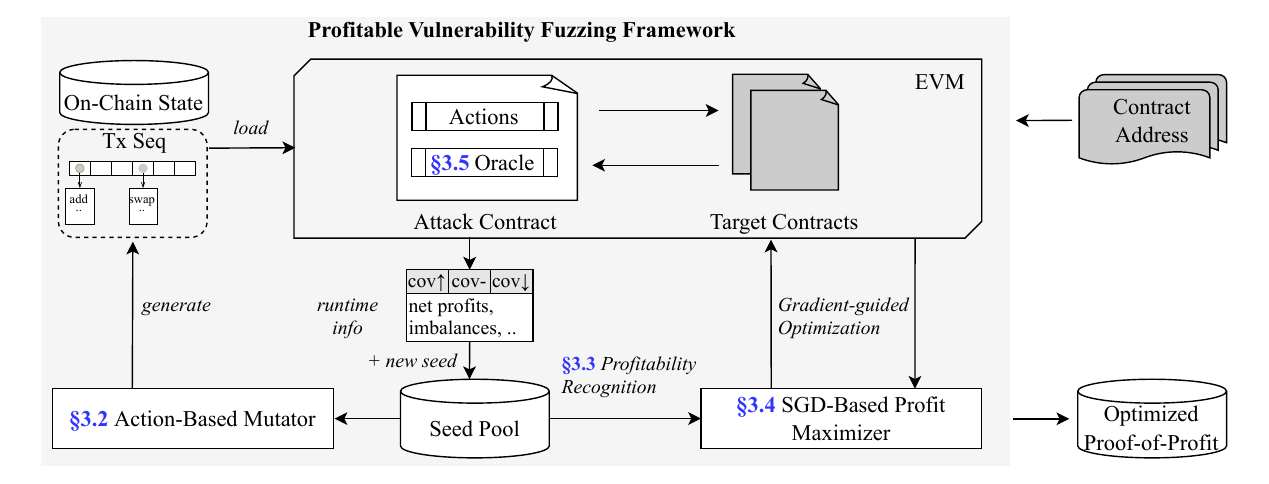}
    \caption{Overall architecture of \tool}
    \label{fig:arch}
\end{figure*}

\subsection{Our Approach}

Figure~\ref{fig:arch} illustrates the overall architecture of \tool, an on-chain fuzzing platform with a profit-centric fuzzing loop. 
Initially, users provide the target contract address, and \tool fetches the code and interfaces of the target contract accordingly and starts fuzzing.
During each iteration of fuzzing, \tool generates several transactions, each of which encodes generated actions and accounting oracles using the multicall pattern\footnote{\url{https://github.com/OpenZeppelin/openzeppelin-contracts/blob/v4.9.5/contracts/utils/Multicall.sol}}, and then calls the attack contract with the transactions.
The lower left part of the figure depicts the general fuzzing loop of \tool, which employs an action-based mutator (Section~\ref{sec:action}) to generate transactions with diverse fund flows.
Throughout the fuzzing process, if any input seed is identified as potentially profitable (Section~\ref{sec:candidate}), it is elevated to the SGD-based profit maximizer (Section~\ref{sec:sgd}) for parameter optimization.
The input that ultimately yields the highest positive profit (Section~\ref{sec:acc}) is returned as the output of the fuzzer.
\compactline

\section{Methodology}\label{sec:method}

\subsection{Problem Modeling \& Approach Overview}

\noindent
\textbf{State, Input, Execution, and Fund Flow} \tab
The initial state of \tool is a world state $S_{n}$ at a given block $n$, and the input of \tool $I$ can be defined as a sequence of transactions $I = [T_{0}, T_{1} ... T_{k-1}]$ and $T_{i}$ could be transaction generated by \tool or an existing pending transaction. A single execution of \tool is $k$ times consecutive states transitions of the initial world state as shown in equation \ref{eq:transit}.

\begin{equation}\label{eq:transit}
S_{n}^{0} \xrightarrow{T_{0}} S_{n}^{1} \xrightarrow{T_{1}} ... \xrightarrow{T_{k-1}} S_{n}^{k}
\end{equation}

During the execution, \tool also collects the fund flows of each transaction and builds the fund flow graph $G(V, E)$ at the end, where the vertices $V$ are defined as the set of all involved addresses and the edges $E$ are defined as the set of token transfers.

\noindent
\textbf{Profit} \tab
\tool features a profit accounting design to compute the assets of attackers in a single token at any state. Given $N(S)$ as the accounting function, we define the profit $P(I)$ of a \tool input $I$ in equation \ref{eq:profit}.

\begin{equation}\label{eq:profit}
P(I) = N(S_{n}^{k}) - N(S_{n} ^ {0})
\end{equation}

\noindent
\textbf{Goal} \tab
When an input $I$ satisfies $P(I) > 0$, \tool reports it as a \textit{proof-of-profit}, which also acts as a trivial fuzzing oracle. The goal of \tool is thus to find a proof-of-profit that yields the maximum profits, as defined in equation \ref{eq:goal}.

\begin{equation}\label{eq:goal}
\max_{I} P(I), \: \textrm{where} \:  P(I) > 0
\end{equation}

\noindent
\textbf{Approach Overview} \tab
Algorithm \ref{alg:loop} outlines the solution \tool{} proposed for finding the optimized proof-of-profit.
It begins by initializing the maximum profit $\mathcal{P}$ and the corresponding input $\mathcal{I}$ to zero and null, respectively.
The main loop continues until a timeout condition is met.
First, an input $I$ is selected from the corpus $C$ using the function \texttt{SelectInput}.
This input is then mutated by \texttt{ActionMutate} to explore variations of the transaction sequence, as detailed in Section~\ref{sec:action}.
The mutated input $I'$ is executed on the current state $S_n$, resulting in a new state $S'_n$ and a corresponding fund flow graph $\mathcal{G}$, as described in Section~\ref{sec:action}.
The profit of this state is then calculated using \texttt{ProfitAccounting}, which evaluates the assets in $S'_n$, as outlined in Section~\ref{sec:candidate}.
The algorithm checks if $I'$ is a potential candidate for a proof-of-profit by using \texttt{IsPoPCandidate} to analyze the graph $\mathcal{G}$ and states $S'_n$ and $S_n$, as explained in Section~\ref{sec:acc}.
If it is, the input and profit are further optimized using the Stochastic Gradient Descent method, as discussed in Section~\ref{sec:sgd}.
If the newly calculated profit $P'$ exceeds the current maximum profit $\mathcal{P}$, the algorithm updates $\mathcal{P}$ and $\mathcal{I}$ with $P'$ and $I'$, respectively, ensuring that the maximal proof-of-profit will be returned.

\subsection{Action-Based Mutator}\label{sec:action}

\begin{algorithm}[tp]
\SetFuncSty{sffamily}

\DontPrintSemicolon
\SetKwInput{KwInput}{Input}                
\SetKwInput{KwOutput}{Output}              
\SetKwFunction{Timeout}{Timeout}
\SetKwFunction{SelectInput}{SelectInput}
\SetKwFunction{ActionMutate}{ActionMutate}
\SetKwFunction{ProfitAccounting}{ProfitAccounting}
\SetKwFunction{Execute}{Execute}
\SetKwFunction{SGD}{SGD}
\SetKwFunction{IsPoPCandidate}{IsPoPCandidate}

\SetKwData{And}{and}

\KwInput{Initial State $\mathcal{S}_{n}$, Corpus $\mathcal{C}$.}
\KwOutput{Proof-of-Profit $\mathcal{I}$ with maximized profits $\mathcal{P}$}

$\mathcal{P}, \mathcal{I} \gets 0, null$\;

\While{$\neg\Timeout()$}{
    $I \gets \SelectInput(\mathcal{C})$\;
    $I^{\prime} \gets \ActionMutate(I)$ \tcp{\small Section \ref{sec:action}}
    \;
    $\mathcal{S}_{n}^{\prime}, \: \mathcal{G}$$ \gets \Execute(I^{\prime}, \mathcal{S}_{n})$ \tcp{\small $\mathcal{G}$ is the fund flow graph.}
    $P^{\prime}$ $\gets \ProfitAccounting(\mathcal{S}_{n}^{\prime})$ \tcp{\small Section \ref{sec:acc}}
    \;

    \If(\tcp*[h]{\small Section \ref{sec:candidate}}){$\IsPoPCandidate(\mathcal{G}, \mathcal{S}_{n}^{\prime}, \mathcal{S}_{n})$} 
    {
        $P^{\prime}, I^{\prime} \gets \SGD(I^{\prime}, \mathcal{S}_{n})$ \tcp{\small Section \ref{sec:sgd}}\label{alg:gradline}
    }
    \;
    
    \If(\label{alg:eval}){$P^{\prime} > 0 \:\And\: P^{\prime} > \mathcal{P}$}
    {
        $\mathcal{P} \gets P^{\prime}$\;
        $\mathcal{I} \gets I^{\prime}$
    }   
}
\caption{Fuzzing loop of \tool}
\label{alg:loop}
\end{algorithm}

\begin{table}[t]
    \centering

    \begin{tabular}{c|c|c}
    \toprule
    Name & Source & \multicolumn{1}{c}{Action}   \\
    \hline
    \grayrow
    A1 & ERC20 & Transfer tokens at a given percentage of current balance. \\
    A2 & Uniswap V2 & Swap tokens using a Uniswap V2 router.  \\
    \grayrow
    A3 & Uniswap V2 & Add or withdraw liquidity to a Uniswap V2 pair. \\
    A4 & Uniswap V2 & Use the low-level Uniswap V2 pair function to do swap or flashloan. \\
    \grayrow
    A5 & Uniswap V2 & Use the low-level Uniswap V2 pair function to mint or burn liquidation. \\ 
    A6 & Uniswap V3 & Use the low-level pool function to borrow flashloan. \\
    \grayrow
    A7 & Uniswap V3 & Swap tokens using a Uniswap V3 router. \\
    A8 & Uniswap V3 & Add or withdraw liquidity to a Uniswap V3 pool. \\
    \grayrow
    A9 & DODO V2 & Borrow flash loan from a DODO V2 pool. \\
    A10 & DODO V2 & Swap tokens using a DODO V2 pool. \\
    \bottomrule
    \end{tabular}
    \caption{Actions extracted from DODO, Uniswap and the ERC20 documents.}
    \label{tab:actions}
\end{table}

Smart contract fuzzing presents a challenging search space where both API sequences and their parameters must be semantically correct to execute transactions effectively.
Existing fuzzers~\cite{smartian,ityfuzz} typically generate transactions at the API level by randomly selecting API sequences and parameters, which often proves inefficient and ineffective for producing valid transactions.
To improve upon this, \tool employs action-based mutation.
In \tool, a transaction $T_{i}$ within input $I$ can either be a call to external contract functions or a high-level DeFi action
~\footnote{Internally represented as a transaction to the attacker contract itself.} 
composed of a series of predefined sub-transactions.
Although new and customized actions can theoretically be introduced continuously, \tool summarizes eight actions that are essential for detecting profitable vulnerabilities in the current smart contract ecosystem.
Table~\ref{tab:actions} provides details on these actions.
One of the actions is the transfer functionality of the ERC20~\cite{erc20} token, while the others are common operations supported by leading decentralized exchanges (DEXs) such as Uniswap~\cite{uniswap, uniswapv3} and DODO~\cite{dodo}.
These actions were chosen due to their significant role in the ecosystem: Uniswap and DODO alone account for over 75\% and 90\% of the weekly trading volumes of ERC20 tokens on Ethereum mainnet (ETH) and Binance Smart Chain (BSC), respectively~\cite{defilama}.
The actions are implemented by rewriting code snippets from documentation samples or unit tests~\cite{uniswapexamples}.
By selecting these prevalent on-chain projects, \tool ensures it targets the most impactful areas for vulnerability detection.
Furthermore, to enhance flexibility and support a broader range of actions, \tool automatically detects actions defined in the attacking contract.
This feature allows for the extension of actions by writing new ones in Solidity without needing modifications to \tool's internal fuzzing implementation.
More guidance on semi-automated extending supported actions with LLM-based methods can be found on our website~\cite{ourwebsite}.

As mentioned, in the generated transaction sequence, each transaction can be either an action or a single invocation of an external function.
When generating a new transaction, \tool applies action-based mutation by randomly deleting actions, mutating action arguments, or inserting new actions.
To better illustrate how actions work, the Listing \ref{lst:action} shows the simplified implementation of A2 action defined by \tool{} in the attacking contract.
By defining such actions in advance, \tool can reduce its search space in two ways:

\begin{itemize}
    \item \textbf{Satisfy API Dependencies} As suggested in the Listing \ref{lst:action}, users shall always call \verb|approve| before doing swaps. By combining the two external calls into a single action, we ensure that the swap between two tokens on UniswapV2 won't revert due to wrong sequence of the two calls.
    \item \textbf{Constrain Parameters} The third parameter of action A2 is not the common absolute balance but instead "percentage", which indicates the portion of the sender's balance. Therefore, \tool will always mutate the "percentage" within $[0, 100]$. Without this constraint, \tool is hard to generate an unsigned integer for the swap to succeed. For instance, if the sender holds 1e30 tokens, the probability of randomly generating a valid balance to transfer is $\frac{1e30}{2^{256} - 1} << 0.0001$, which is almost impossible.
\end{itemize}

For simplicity, \tool does not make any explicit assumptions about the inter-action dependencies, meaning any two actions can be combined freely during mutation.

\subsection{Profitability Recognition}\label{sec:candidate}

Instead of maximizing profits for every seed, \tool evaluates seeds by analyzing their fund flow graphs and the resulting state of inputs.
Only inputs with high profit potential, termed \textit{Proof-of-Profit Candidates}, undergo the SGD process.
\tool uses the following criteria to identify these candidates:

\begin{itemize}
    \item \textbf{Positive Profits}: Inputs that already demonstrate positive profits before the profit maximization process are classified as proof-of-profit candidates.
    \item \textbf{Imbalanced Uniswap Pairs}: If a transaction $T_{i}$ causes the reserves of Uniswap pairs to be unequal to their expected balances, indicating potentially sub-optimal exchange rates \cite{uniswap}.
    \item \textbf{Unconditional Token Gain}: The fund flow graph $G$ of a transaction $T_{i}$ shows that attackers can gain tokens without incurring any cost, potentially inflating the token supply.
    \item \textbf{Unconditional Token Burn}: The fund flow graph $G$ of a transaction $T_{i}$ reveals that attackers can burn tokens of other holders without costs, which could deflate the token supply.
\end{itemize}

The intuition of proof-of-profit candidates is that if an input fits into any of the rules above, it could probably cause abnormal price fluctuation, which usually implies the possibility of gaining further profits.
Even if the profit of the candidates is negative, \tool will prioritize the candidates when selecting inputs and try to increase its profits from negative to positive by gradients.
More running details of this process can be found in Section~\ref{sec:rq2}.

\subsection{Stochastic Gradient Descent Based Profit Maximizer}\label{sec:sgd}

Once a proof-of-profit candidate $I$ is determined to go through the SGD process, \tool will identify and record the single transactions $T_{i}$ causing the abnormal price fluctuation.
\tool models the optimization of the variables inside these transactions as the problem of finding the maximal value of $f(x_1, x_2, ..)$, where $f$ represents the execution of the transaction sequence and $x_i$ denotes all fuzzer controllable variables of the transactions.
To boost the search of maximal value, \tool{} adopts stochastic gradient descent (SGD) to iteratively update the variables $x_i$ in the transactions.
The variables of the transactions contains:

\begin{itemize}
    \item \textbf{Integers in Parameters} This includes any signed or unsigned parameters in the input of calling contracts functions.
    \item \textbf{Values of Transactions} The transaction value, representing the amount of ether sent by the caller, is deemed a variable.
    \item \textbf{Repeats} Each transaction's repeat times are considered variables.
\end{itemize}

Based on these variables and the profit function defined in Equation~\ref{eq:profit}, \tool{} calculates its gradients $\bm{g}$ by adding a small enough value $\delta$\footnote{$\delta$ can be negative because the gradients $\frac{\partial P(\bm{x})}{\partial x_{i}}$ could be not continuous.} to $x_{i}$ and computing the profits difference as shown in Equation~\ref{eq:gradprofit}, where $\bm{g}$ denotes the gradients and $\bm{v}_{i}$ is a unit vector that only the position of $x_{i}$ is 1.

\begin{equation}\label{eq:gradprofit}
    \bm{g} = \frac{\partial P(I)}{\partial x_{i}} = \frac{\partial P(\bm{x}) } {\partial x_{i}} = \frac{P(\bm{x} + \delta \bm{v}_{i}) - P(\bm{x})}{\delta}
\end{equation}

\noindent
\textbf{Gradient Selection} \tab \tool features a biased selection of gradients instead of randomly selecting gradients to descent. \tool will favor $x_{i}$ if its corresponding transaction $T_{i}$ is a direct cause of proof-of-profit candidate as defined in section \ref{sec:candidate}. This helps \tool concentrate its fuzzing power on possibly increasing the profit potential.

\noindent
\textbf{Step Optimization} \tab In theory, we can maximize the profit of any input by substituting $\bm{x}$ with $\bm{x} + \delta \bm{v}_{i}$ and loop until all gradients are zero.
However, it is not feasible in practice because it takes too many iterations in some cases, given the integers are 256 bits.
Therefore, picking an appropriate step $\bm{\alpha}$ is crucial for SGD to converge to the maximum profits in an acceptable period.
To solve this challenge, \tool heuristically features a step selection algorithm in Equation \ref{eq:step}, where $g_{i, j}$ and $\alpha_{i, j}$ denotes the gradient value and corresponding step of independent variable $x_{i}$ in $j$-th SGD loop.
$R(n)$ means generating a random integer below $n$.

\begin{equation}\label{eq:step}
    \alpha_{i, j+1}=
    \left\{ \begin{array}{ll}
        p \alpha_{i, j} & g_{i, j} \cdot g_{i, j+1} >0 \: \text{and $x_{i}$ was taken from parameters or values} \\
        \alpha_{i, j} + R(n) & g_{i, j} \cdot g_{i, j+1} >0 \: \text{and $x_{i}$ was taken from repeat times} \\
        q \alpha_{i, j} &  g_{i, j}  \cdot g_{i, j+1} < 0 \\
        0 & g_{i, j + 1} = 0
    \end{array} \right.
\end{equation}

Regarding the parameter $p, q, n$, the basic intuition of the Equation \ref{eq:step} is that if the gradient direction in the $j+1$ loop is the same as the previous loop, we will take a bigger step, expecting a faster convergence. With the opposite gradient direction, we will take a reduced step with an opposite direction to avoid diverging from the maximum point.
The step of $x_{i}$ from repetitions are separately calculated with smaller steps to avoid gradient vanishing because of revert.
The three parameters strongly correlate with properties of the profit function $P$, which doesn't have an analytic form.
Therefore, there is no general optimal step parameter. However, as long as the parameters remain in reasonable values, \textit{i.e.}, $p-1$ is not too close to 0, $p$ is not excessively large, $p>-q$ and $n \leq 9$, the impact of parameters' selection is quite limitted because we actually adopt a variant of binary search with lookback mechanism.
\tool empirically uses $p = \frac{3}{2}, q = - \frac{1}{3}, n = 5$ for all experiments of this paper.

\noindent
\textbf{Gradient Explosion and Vanishing} \tab
Unlike many common optimizers, the gradients $\bm{g}$ are intentionally excluded from step computations to mitigate frequent gradient explosions.
Additionally, gradient vanishing is often encountered during the SGD process because the profit function $P$ is undefined when the transaction sequence reverts, resulting in a segmented domain for $P$.
In such cases, \tool attempts to locate the boundary of the current segment of $P$'s domain.
Specifically, when gradient explosion or vanishing occurs, the value $x_{i, j+1}$ is reverted to $x_{i, j}$, and a smaller step is tried to identify the boundary of $x_{i}$.

\noindent
\textbf{Local Maximum} \tab SGD is known to suffer from converging to a local maximum. \tool mitigates this issue by only mutating the arguments of the inputs to produce new inputs and go through SGD process again. This effectively selects another starting point in the problem space of SGD in hoping of finding the global maximum.

\subsection{On-Chain Profit Accounting Oracle}\label{sec:acc}

During attack generation, the assets attacker holds usually are not in a single token. Since \tool needs to compare the profits of two inputs, as shown in the line \ref{alg:eval} of Algorithm \ref{alg:loop}, it is crucial to have a mechanism to account for all tokens to a single pricing token to compute profits.
A trivial solution could be fetching the prices of all tokens and computing the whole value in the pricing token. However, this is not precise enough for on-chain fuzzing, leading to false positives when generating attacks. For example, some tokens may lack the liquidity to swap or be locked due to constraints, and thus, the price may not reflect the real value.
\compactline

To solve this challenge, \tool features a unique on-chain accounting process, which eventually is implemented based on the actions defined in Table~\ref{tab:actions} in Solidity.
During the fuzzing process, \tool will closely collect the involved assets by monitoring the fund flow graph.
Once execution is over, \tool will perform the on-chain accounting by the following steps.

\begin{itemize}
    \item  \textbf{Withdraw All Assets} The on-chain assets can exist in various forms, like ERC20 tokens, the liquidity provided to Uniswap V2/V3 pools, or the balance of attacker-controlled accounts in native currency. \tool will first withdraw all related assets to ERC20 tokens.
    \item  \textbf{Swap via Decentralized Exchanges} \tool will compare different quotes across various decentralized exchanges and select the best one to swap all tokens to the pricing token, usually some mainstream tokens like \textit{USDT} and \textit{ETH}. The output of the on-chain accounting is the final amount of the pricing token.
\end{itemize}

This ensures that \tool can quote the actual value of all assets in a single mainstream token, which we deem concrete proof-of-profits.

\noindent
\textbf{Pricing Token Mutation} \tab It is possible that some assets don't have decentralized exchanges to swap, e.g., the liquidity between the assets and the pricing token is zero.
To overcome this challenge, the mutator of \tool will also mutate the pricing token so that \tool can support trading as many assets as possible.
Since the profits in different pricing tokens can not be compared directly, \tool will maintain and output maximum proof-of-profit in different pricing tokens. 

\section{Evaluation}\label{sec:eval}


\noindent
\textbf{Evaluation Questions} \tab
In this section, we aim to answer the following research questions.

\begin{itemize}

    \item \textbf{RQ1} \tab
    How does \tool perform compared with the state-of-the-art fuzzing solutions?
    
    \item \textbf{RQ2} \tab
    How do the key designs of \tool{} impact its overall effectiveness?
    
    \item \textbf{RQ3} \tab
    How does \tool{} perform in detecting unknown smart contract vulnerabilities?
    
\end{itemize}

\noindent
\textbf{Implementation} \tab
\tool is written from scratch with 10,106 lines of Rust code and 822 lines of Solidity code.
We use the revm~\cite{revm} as our EVM emulator and libafl~\cite{libafl} to build our fuzzer.
There are roughly 300 lines of Solidity code to define actions and 2,000 lines of Rust code to identify and do action-based mutation automatically.
\tool is highly flexible and modularized and easy to migrate to other networks since we abstract over the execution engines and assets accounting.
In addition, we also implemented four variants of \tool for the ablation study to answer RQ2:

\begin{itemize}
    \item \toolnoaction: In this variant, our mutator will no longer pick the actions defined in Table \ref{tab:actions} but instead always select a random external contract function to call.
    \item \toolnoabnormal: Only positive profit candidates will go through the profit maximizer. Non-positive profit candidates will no longer be considered as proof-of-profit candidates.
    \item \toolnoaccount: The on-chain accounting function $N$ defined in Equation \ref{eq:profit} is substituted with a trivial function $N^{\prime}$ that just returns the balance of the pricing token instead of withdrawing all assets of attackers and trading them to compute the actual value.
    \item \toolnograd: The gradient function defined in the line \ref{alg:gradline} of Algorithm \ref{alg:loop} is removed in this variant.
\end{itemize}

\noindent
\textbf{Baseline} \tab To answer RQ1, we evaluated \tool against two baselines: \ityfuzz~\cite{ityfuzz}\footnote{Commit hash: 8fdcdf7851c7e66ec8e435d555ef4a4fd217a3b6} and on-chain real-world attacking transactions of the incidents collected in our datasets.
\ityfuzz is picked because it is the well-maintained, state-of-art on-chain fuzzer.
The real-world attacking transactions are usually highly optimized because hungry on-chain searchers tend to instantly fill any residual profits due to the "Dark Forest" nature of the blockchain~\cite{darkforest}.
We further excluded fuzzers that only support finding profitable vulnerabilities in native currency because of the popularity of ERC20 tokens. It will take non-trivial extra development efforts to modify them to support detecting ERC20-related profitable vulnerabilities.
Specifically, we once evaluated ConFuzzius~\cite{confuzzius} and EF/CF~\cite{efcf} with the same experiment setup on our dataset.
However, aligning with recent research work~\cite{demystifying}, none of them found any proof-of-profit due to lacking concepts modeling like tokens and liquidty, which our on-chain profit accounting oracle does.


\begin{table}[ht]
    \centering
\begin{tabular}{l|ll|l|llll}
    \toprule
    \multirow{2}{*}{Targets} & \multirow{2}{*}{\ityfuzz} & \multirow{2}{*}{Real-World} & \multirow{2}{*}{\tool} & \multicolumn{4}{c}{\textsc{\tool-NO}} \\
    \cline{5-8}
    & & & & \textsc{- ACT}  & \textsc{- CDT} & \textsc{- ACC} & \textsc{- GRD}\\ 
    \midrule
    \grayrow
    uranium & - &  40,814,877.9 &  17,013,205.4 & 97.0\% & 97.0\% & - &  2.7\%  \\
    zeed & - & 1,042,284.8  &  0.03 & 49.2\% & 20.0\% & 40.0\% & 18.0\% \\
    \grayrow
    hackdao   & -  &  56,973.5 &  \textbf{64,565.4} & - & \textbf{100.0\%} & - & 80.9\% \\
    shadowfi    & - &  299,006.4  & 298,858.8 & - & - & - & -\\
    \grayrow
    pltd    & - &  24,493.0  &  \textbf{24,497.9} & - & \textbf{100\%} & \textbf{100\%} & 96.8\% \\
    hpay    & - &  31,415.7  &  1.5& 100.0\% & 100\% & 98.7\% & 100\% \\
    \grayrow
    health    & - &  4,539.8 &  \textbf{8,742.5}& - & - & 14.6\% & - \\
    bego    & 3,230.0 &  3,235.2 &  \textbf{3,237.2}& \textbf{100\%} & \textbf{100\%} & \textbf{100\%} & \textbf{100\%} \\
    \grayrow
    seaman & 17.7  &  7,775.6 &  1,260.8 & - & 0.04\% & - & 44.8\%  \\
    mbc    & 1,000.0 & 5,904.4 &  3,443.9 & - & 25.74\% & - & 20.0\% \\
    \grayrow
    rfb    & FP   &  3,526.2 &  \textbf{3,796.2}& - & \textbf{100\%} & - & 57.1\% \\
    aes    & 531.9 &  61,608.0 &  \textbf{63,394.4}& - & - & - & -  \\
    \grayrow
    dfs    & - &  1,458.1 &  \textbf{16,700.3}& - & \textbf{96.0\%} & \textbf{100\%} & \textbf{58.2\%}  \\
    upswing   & 246.0  &  590.1  &  580.6& - & 66.7\% & - & 50.0\% \\
    \grayrow
    thoreumfinance   & -&  1,776.7  &  145.8& - & 0.02\% & 0.01\% & 0.02\%  \\
    bevo    & 8,712.1 &  44,377.3 &  10,270.4& <0.01\% & 98.0\% & <0.01\% & <0.01\% \\
    \grayrow
    safemoon   & -  & 8,574,004.4  &  10,492.4& 100.0\% & 100.0\% & 100\% & 100\% \\
    swapos   & - &  278,903.0  &  276,306.7& 100.0\% & 100.0\% & 100.0\% & 80.0\% \\
    \grayrow
    olife    & - &  9,966.9 &  \textbf{10,334.3}& - & - & - & -  \\
    axioma    & 21.3 &  6,904.9 &  6,902.4& 100.0\% & 100.0\% & 100.0\% & 100.0\%\\
    \grayrow
    melo   & \textbf{92,051.4}&  90,607.3 &  \textbf{92,303.0} & \textbf{100.0\%} & \textbf{100.0\%} & \textbf{100.0\%} & \textbf{100.0\%} \\
    fapen    & 621.4  &  635.8 &  \textbf{639.8} & \textbf{100.0\%} & \textbf{100.0\%} & \textbf{100.0\%} & \textbf{100.0\%} \\
    \grayrow
    cellframe    & FP &  75,208.6 &  192.4  & - & - & 70.0\% & - \\
    depusdt    & - &  69,786.6 &  37,791.3 & - & - & -  \\
    \grayrow
    bunn   & FP&  12,969.8&  4.2 & 8.3\% & - & - & - \\
    bamboo   & 42.0  &  50,210.1 &  34,491.3 & - & 2.6\% &  2.6\% & 3.0\% \\
    \grayrow
    sut    & FP &  8,033.7 &  \textbf{9,713.8} & \textbf{100.0\%} & \textbf{82.4\%} & - & 62.5\% \\
    uwerx    & - &  321,442.1 &  321,442.1 & - & 100.0\% & 100.0\% & 99.2\% \\
    \grayrow
    gss    & FP &  24,883.4 &  \textbf{25,000.9}& - & 
    \textbf{100.0\%}& - & 99.3\%  \\
    \hline
    Sum   & 106,474 &  51,927,399.2 &  18,338,315.7 & 92.2\% & 94.7\% & 4.1\% & 6.7\% \\
    \grayrow
    FP & 16\tablefootnote{The 11 targets: \textit{valuedefi}, \textit{pancakehunny}, \textit{roi}, \textit{tinu}, \textit{lw}, \textit{selltoken}, \textit{sellc03}, \textit{cfc}, \textit{shido}, \textit{lusd}, \textit{newfi}, are omitted in the table because only \ityfuzz finds \textbf{False Positives} on them and \tool reports nothing.}  & -&  \textbf{0} & \textbf{0} & \textbf{0} & \textbf{0} & \textbf{0} \\
    \bottomrule
    \end{tabular}
    \caption{The maximum profits gained by \tool, \ityfuzz and real-world attacking transactions. The projects where both fuzzers fail to find any proof-of-profit or only \ityfuzz finds false positives are omitted. The profits higher than those gained by real-world attacking transactions are highlighted in bold. "FP" refers to "False Postive". The profits of ablation variants are represented in the percentage of the profits \tool gained.
    }
    \label{tab:profits}
\end{table}


\noindent
\textbf{Dataset Construction} \tab To evaluate \tool, we extended the dataset \ityfuzz used~\cite{ityfuzz} by another half year following the same criteria, from the same data source, DefiHackLabs~\cite{defihacklabs}.
Our new dataset consists of 61 open-source Web3 projects relating to Uniswap and ERC20 tokens exploited on the Ethereum mainnet (ETH) or Binance Smart Chain (BSC) in the past two years, including a variety of bug types.
The loss amount ranges from hundreds to millions of dollars, with an average of \$1.1 million per project.
We extracted all relevant participant addresses (except the attacker and attacker-controlled contracts) from the attack and used the block number just before the attack to fork the specific chain.
The full dataset can be found on our website~\cite{ourwebsite}.

\noindent
\textbf{Experiment Setup} \tab
We evaluated all fuzzers on two machines, both running Ubuntu 22.04 with 128 cores and 1TB memory, and saved all intermediate data in tmpfs to minimize IO overhead.
Because of the nature of on-chain fuzzing, we need quick access to the full blockchain history, and thus, we spun up archive nodes on a separate machine with 32 cores and 128GB memory to avoid interfering with fuzzers.
We ran all fuzzers with a 12-hour timeout and 2GB memory limit and repeated five times.
If any fuzzer ran out of memory and got killed, we would restart the fuzzer, and the longest campaign that reports the maximum profit was selected at the end of the evaluation.
All fuzzer instances were wrapped in docker containers and bound to a single core.

\subsection{\textbf{RQ1} State-of-the-Art Comparison}\label{sec:rq1}

To address \textbf{RQ1}, we compiled a subset of targets from our dataset, specifically those where at least one fuzzer identified a true finding, as shown in Table \ref{tab:profits}.
Recognizing that some fuzzers may produce false positives due to design or implementation issues, we verified the proof-of-profit generated by both fuzzers.
This was done by forking the blockchain at the relevant height and manually replaying transactions to eliminate any false positives.

\noindent
\textbf{Overall Results} \tab \tool gains over 18 million profits by successfully generating concrete exploits for 29 targets and outperforms \ityfuzz on all evaluated targets without any false positives.
We would like to highlight that \tool can even gain more profits than highly optimized real-world attacking transactions on 12 targets (from 1.01 to 11.45 times more). 
All these statistics prove the effectiveness of \tool.

\noindent
\textbf{Compared with Real-World Attacking Transactions} \tab
The following provides an understanding of why and how \tool outperforms or underperforms real-world attacking transactions.

\textbf{\ding{172} Analysis on Targets Where \tool Outperforms Real-World Attacks} \tab
To understand why \tool can even outperform on-chain attacking transactions, we manually analyzed the transaction sequences generated by \tool and the attacking transactions and categorized the targets in two reasons.

\begin{itemize}
    \item \textbf{More Optimal Parameters} The parameters adjusted by the gradient of \tool are more optimal than attackers' transactions, though the transaction sequences are exactly the same. This includes \textit{bego}, \textit{melo}, and \textit{gss}.
    \item \textbf{Better Transaction Sequence} \tool generates a better transaction sequence to gain more profits. This includes all the other targets.
\end{itemize}

\begin{lstlisting}[float=t, language=Python, caption=\textit{bego} vulernerability, label={lst:bego}, escapechar=|]
function _mint(address account, uint256 amount) internal {
    require(account != address(0), "BEP20: mint to the zero address");
    // ...
    _balances[account] = _balances[account].add(amount);|\label{line:bego}|
    // ...
}
\end{lstlisting}

We will show the case of \textit{bego}\footnote{Address:  \href{https://bscscan.com/address/0xc342774492b54ce5f8ac662113ed702fc1b34972}{0xc342774492b54ce5f8ac662113ed702fc1b34972}} to illustrate the first point as shown in the Listing \ref{lst:bego}.
Its \verb|mint| function lacks sufficient permission checks and allows anyone to mint \textit{bego} tokens without any cost, which can immediately be captured as a proof-of-profit candidate and sent to the gradient process. Obviously, the final profit function $P$ increases monotonically with respect to the \verb|amount| argument of the \verb|mint| function. Therefore, \tool can easily determine the best \verb|amount| as \verb|3634607800974379339971347430454067| for the maximum profits draining all the liquidity while the on-chain attacking transaction simply uses \verb|1e30| with the same transaction sequence and thus causing the slight difference in profits.

Another case is \textit{sut}\footnote{Address: \href{https://bscscan.com/address/0xf075c5c7ba59208c0b9c41afccd1f60da9ec9c37}{0xf075c5c7ba59208c0b9c41afccd1f60da9ec9c37}}, as shown in Listing \ref{lst:sut}.
The line \ref{line:sut} uses \verb|tokenPrice| as the token's price to calculate the return tokens, but the price is fixed and too low. The on-chain attacking transaction exploited this by an arbitrage flavor: buying some tokens via the vulnerable function and selling them elsewhere with a much better price.
\compactline

\tool exploits this differently. When \tool calls this function with $0 < \_numberOfTokens < 1,000,000,000,000,000,000$, the division of the line \ref{line:sut} will be zero because of integer division, and  \verb|_bnbValue| is also zero. Thus, \tool gains tokens without any payment, which is "Unconditional Token Gain" as defined in section \ref{sec:candidate}. By going through SGD, \tool repeatedly calls this function to gain tokens until draining all the liquidity. This transaction sequence is more optimal than the on-chain attacking transaction because it doesn't cost anything and doesn't need a flash loan to save potential fees further.
\compactline


\textbf{\ding{173} Analysis on Targets \tool Failed to Exploit} \tab Although \tool proves its value by yielding a large amount of profits with concrete proof-of-profits, there are still a few targets where \tool fails to find any positive profits. We also manually analyze these cases and the attacking transactions to understand the failure reasons.

\begin{itemize}
    \item \textbf{Hard Constraints} The first reason for the false negatives is the hard constraints, including magic numbers, addresses, ABI decoding for bytes and block or timestamp constraints. 17 cases fall in this category. Employing symbolic execution or dynamic dataflow analysis or extracting more actions automatically shall mitigate the limitations.
    \item \textbf{Inter-Action Dependency} For instance, the staking pool related functions expect the previous returned staking ID as its argument. This includes 6 cases.
    \item \textbf{Limited Computation Resources} Though theoretically solvable, \tool fails to exploit the left nine targets within the given CPU cores and fuzzing duration.
\end{itemize}


\begin{lstlisting}[float=t, language=Python, caption=\textit{sut} vulernerability, label={lst:sut}, escapechar=|]
function buyTokens(uint256 _numberOfTokens) public payable {
    uint256 _bnbvalue = (_numberOfTokens/1000000000000000000)*tokenPrice; |\label{line:sut}|
    require(msg.value >= _bnbvalue);
    // ...
    require(
    tokenContract.transfer(msg.sender, _numberOfTokens),"Transfer failed");
    // ...
}
\end{lstlisting}

\noindent
\textbf{Compared with \ityfuzz} \tab \ityfuzz can only generate profitable attacks on ten targets, at a sum of 100k profits while \tool outperforms \ityfuzz by finding more profits on all the ten targets and yielding 172 times more total profits.

\textbf{\ding{172} Proof-of-Profit Candidates Identification \& Optimization} \tab
To understand the reason why \tool outperforms \ityfuzz, we collected the proof-of-profit candidates identified by both fuzzers as shown in Table \ref{tbl:candidate}.
Of all 61 targets, \tool manages to identify proof-of-profit candidates on 45 targets, significantly more than \ityfuzz does.
In addition, \tool will utilize the information gained from the candidates during the input selection and SGD while \ityfuzz reports the vulnerabilities and stops further exploration.
This helps \tool in exploiting the profit potential of proof-of-profit candidates.
\compactline

\begin{table}[t]
    \centering
\begin{tabular}{l|l|l}
    \toprule
    \multicolumn{1}{c|}{Proof-of-Profit Candidate Type} & \tool & \ityfuzz \\
    \hline
    Imbalanced Uniswap Pairs & 37 & 10 \\
    \grayrow
    Unconditional Tokens Gain & 36 & - \\
    Unconditional Tokens Burn & 11 & - \\
    \grayrow
    Positive Profits & 22 & 10 \\
    \hline
    Total & 45 & 19 \\
    \bottomrule
\end{tabular}
\caption{Breakdown of proof-of-profit candidates. \ityfuzz only implemented "Imbalanced Uniswap Pairs" and "Positive Profits". Some targets may satisfy more than 1 rule.}
\label{tbl:candidate}
\end{table}

\textbf{\ding{173} False Positive Analysis} \tab On-chain arm-race is highly automated, and thus, one of the goals of \tool is to reduce false positives as much as possible because false positives will take extra manual efforts for further validation.
Of all the targets we evaluated, \tool has no false positives while \ityfuzz has 16 false positives.
The reasons for the false positives of \ityfuzz are illegal or unreachable EVM states because of hijacking EVM execution or error during off-chain accounting emulation\footnote{Detailed false positive analysis of \ityfuzz will be available at our website~\cite{ourwebsite}.}.
For instance, transactions that shall revert could still be committed during \ityfuzz's emulation, leading to false positives. In contrast, \tool is equipped with on-chain accounting and ensures the profits gained are concrete and reproducible by leaving EVM execution intact.

\begin{lstlisting}[float=t, language=Python, caption=The code snippet of critical lines of \textit{dfs}, label={lst:dfs}, escapechar=|]
function _transfer(address from, address to, uint256 amount ) internal {
    if (to == address(pair) || from == address(pair) ) { 
        if (takeFee && !exclusiveFromFee[from]) {
            fee = amount.mul(rate).div(1000);
            _balance[from] = _balance[from].sub(amount).sub(fee); |\label{line:dfs}|
            _balance[destroyAddress] = _balance[destroyAddress].add(fee);
            emit Transfer(from, destroyAddress, fee);
        }
    }
}
\end{lstlisting}

\textbf{\ding{174} Throughput and Coverage} \tab We collected the instruction coverage and transaction throughput of \tool and \ityfuzz following previous work~\cite{arewe, ityfuzz}.
\tool averagely achieves 3,178 transactions per second and 66.2\% instruction coverage while \ityfuzz only achieves 1,729 transactions per second and 54.0\% instruction coverage. 50 out of 61 projects reach the maximum coverage in first 6 hours while the rest in 12 hours.
\ityfuzz has a lower transaction throughput mainly because of the overhead incurred by taking huge amounts of snapshots due to the complex state space of on-chain fuzzing.
We also fetched the source code mappings from block explorers and analyze the line coverage of the targets that only \tool finds proof-of-profits. It turns out that 8 out of 14 targets\footnote{Six targets do not have source mapping.} have higher coverage and reach the lines \ityfuzz fails to cover due to our action-based mutation and profits-targeted fuzzing strategies.
For instance, We use \textit{dfs}\footnote{Address: \href{https://bscscan.com/address/0x2B806e6D78D8111dd09C58943B9855910baDe005}{0x2B806e6D78D8111dd09C58943B9855910baDe005}} as a case study as shown in Listing \ref{lst:dfs}. \textit{dfs} takes a fee on selling tokens but asks users to calculate the fee in advance, i.e. users can not sell 100 tokens even if users hold exactly 100 tokens. Our action-based mutation supports mutating the ratio of transferring and selling tokens, as detailed in the Section \ref{sec:action}. Thus, we can have higher code coverage by finishing the swapping successfully while \ityfuzz always reverts at Line \ref{line:dfs}. The high throughput and coverage ensure \tool is competitive for on-chain auditing arm-race.


\subsection{\textbf{RQ2} Ablation Study} \label{sec:rq2}


To answer RQ2, we evaluated \toolnoaction, \toolnoaccount, \toolnoabnormal, and \toolnograd on the same datasets with the same experiment setup as shown in Table \ref{tab:profits}.

\noindent
\textbf{Action-Based Mutation} \tab Without the action-based mutation, \toolnoaction can only find profits on half of the targets compared to \tool.
This is because it is much harder to satisfy the API dependencies purely based on random API combination: mostly the generated transactions will eventually be reverted in execution, greatly harming efficiency.

\noindent
\textbf{On-Chain Accounting} \tab \toolnoaccount behaves much worse than \tool by exploiting 13 less targets.
The first reason is that some inputs could have already exploited vulnerabilities, but without accounting, \toolnoabnormal fails to determine they are profitable attacks.
In addition, the lack of withdrawal of all assets also affects the SGD process, which heavily relies on the profit function to report profits, as targets like \textit{health} and \textit{bevo} suggest.

\begin{figure}[t]
    \centering
    \includegraphics[width=\linewidth, trim=0 20 0 0]{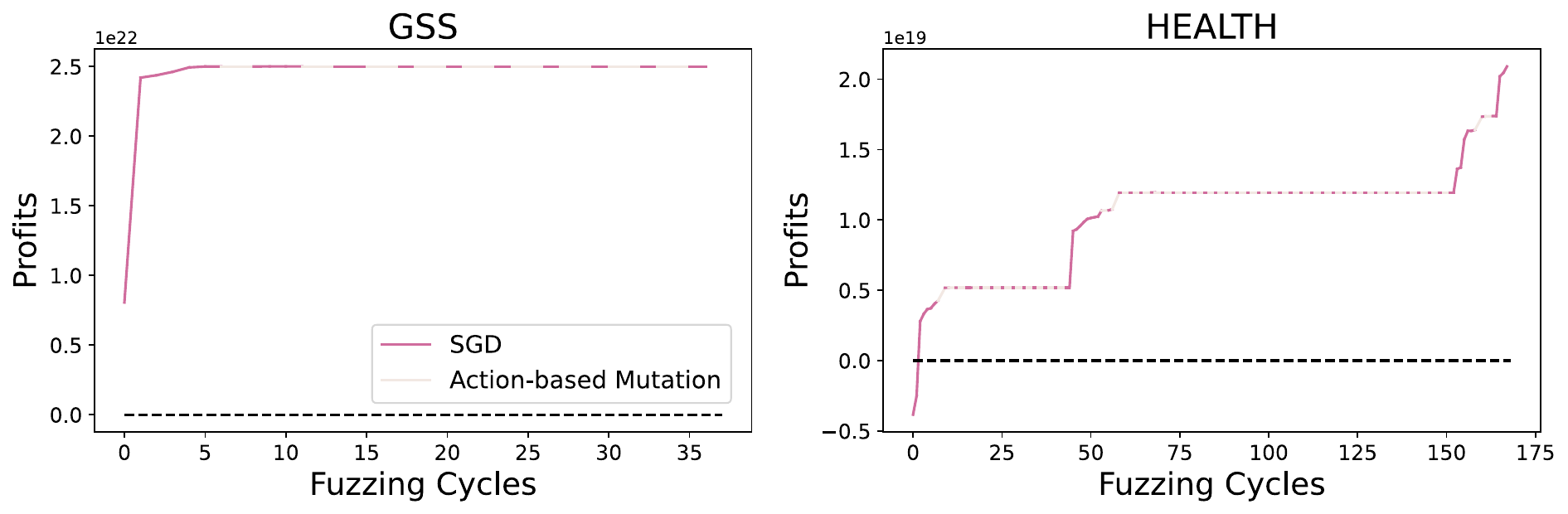}
    \caption{Representative graph of profits increasing while doing SGD. Both graphs are strictly monotonous.}
    \label{fig:gradgraph}
\end{figure}

\noindent
\textbf{Proot-of-Profit Candidate and SGD} \tab \toolnoabnormal and \toolnograd have a similar performance by missing seven targets compared with \tool because both variants do not exploit the profit potential of the inputs with negative inputs.
To illustrate this process better, we chose two representative cases \textit{gss} and \textit{health} from RQ1 to plot how the maximum profits change with the fuzzing campaign going in Figure \ref{fig:gradgraph}.
The \textit{gss} case is typical in that SGD starts from a proof-of-profit and maximizes the profits until converging.
In contrast, \textit{health} case illustrates how SGD exploits the profit potential of inputs with negative profits.
In both graphs, the action-based mutation and SGD contribute to the profit increase alternatively. On average, the time distribution of the action-based mutation exploration stage and SGD stage is 52:48 with the specific ratio depending on the number of proof-of-profit candidates. For projects like \textit{gss} and \textit{health} where proof-of-profits candidates are continously found, \tool could spend siginificant amount of time on the SGD stage while for projects with few or no proof-of-profit canditates, SGD stage takes much less time. For both stages, EVM emulation accounts for almost all ($\geq 99\%$) computation resources. \footnote{Figures of the time distributions, profiles and profit increasing contributions of all targets can be found on our website~\cite{ourwebsite}.}

\subsection{\textbf{RQ3} Real-World Application}\label{sec:rq3}

To answer \textbf{RQ3}, we applied \tool on four real-world projects on the prevailing Web3 auditing contest platforms, Secure3~\cite{secure3}.
In total, \tool succeeds in generating concrete proof-of-profits for six vulnerabilities across the four projects without false positives.
We would like to highlight that five of the vulnerabilities are evaluated as high severity because of the possibility of directly losing funds.
By the time of writing this paper, the vulnerabilities had been fixed already before deployment, and we were awarded bounties of more than \$2,500.
To illustrate the application of \tool on real-world projects, we will showcase two of the vulnerabilities \tool found in the rest of this section.

\noindent
\textbf{Setup Process} \tab Unlike on-chain targets, projects for auditing are usually not deployed yet.
To setup \tool, auditors need to spin up a local fork and run \tool before deploying the projects.
\tool will closely watch new transactions on the fork and start fuzzing on any involved addresses in parallel.
It is worth mentioning that \tool supports both front-running and back-running transactions because \tool can start fuzzing from any given state.
Then, auditors deploy the projects' contracts by sending several transactions and collect \tool's output with a one-hour timeout.

\begin{lstlisting}[float=t, language=Python, caption=The vulnerable \textit{burn} implementation of \vulna., label={lst:burn}, escapechar=|]
function burn(address from, uint256 value) public {
    super._burn(from, value);
}
\end{lstlisting}

\noindent
\textbf{Project \vulna} \tab The project \vulna contains a vulnerability that anyone can burn other users' tokens as shown in Listing \ref{lst:burn}.
This vulnerability is trivially caught and exploited by \tool due to matching "Unconditional Token Burn" as defined in Section \ref{sec:candidate}.
\tool can gain and maximize profit by burning tokens of the pools and manipulating the price. 

\noindent
\textbf{Project \vulnb} \tab  The project \vulnb consists of several components, and there are two contracts related to the vulnerability, \verb|Vault| and \verb|Router| as shown in Listing \ref{lst:vault} and \ref{lst:router} respectively.
\verb|Vault| acts like a common token vault: receives tokens from users and mints shares in return, except supporting investing in multiple token types as shown in line \ref{line:mut}.
Similar to Uniswap, the interfaces of \verb|Vault| are not intended for direct usage, and thus \verb|Router| wraps the call to \verb|Vault| and handles token transfers afterward.

\begin{lstlisting}[float=t, language=Python, caption=Simplified \texttt{Vault::mint} snippet of \vulnb, label={lst:vault}, escapechar=|]
function mint(address sender, address token, ... uint256 amount) {
    // ...
    uint shares = _shares[token];|\label{line:mut}|
    uint balance = IERC20(token).balanceOf(address(this));|\label{line:bal}|
    // ...
    uint out;
    if (shares == 0) { |\label{line:zeroshares}|
        out = amount; |\label{line:first}|
    } else {
        out = (amount * shares) / balance; |\label{line:user}|
    }
    _shares[token] = shares + out;
    _mint(sender, ..., shares);
}
\end{lstlisting}

\begin{lstlisting}[float=t, language=Python, caption=Simplified \texttt{Router::buyLongToken} snippet of \vulnb, label={lst:router}, escapechar=|]
function buyLongToken(uint256 amount) {
    // ...
    _vault.mint(msg.sender, _token, ..., amount);
    // ...
    _token.transferFrom(msg.sender, _vault, amount); |\label{line:transfer}|
}
\end{lstlisting}

The vulnerability roots in the implementation of \verb|Vault::mint| shown in Listing \ref{lst:vault}.
Assume \verb|Vault| is empty, and the first user tries to deposit 10 tokens by calling \verb|Router::buyLongToken|. An attacker could front-run his transaction with two transactions.

\begin{itemize}
    \item \textbf{Calling \texttt{Router::buyLongToken} with \texttt{amount=1}}: Because \verb|Vault| is empty, the condition at line \ref{line:zeroshares} is true and attacker become the sole shares holder of \verb|Vault|.
    \item \textbf{Transfer 10 \texttt{\_token} to \texttt{Vault}}: This essentially a donation to \verb|Vault| but the \verb|balance| defined in line \ref{line:bal} will become 11.
\end{itemize}

After the attack has finished his transactions, the user will get his shares by the line \ref{line:user} because the \verb|Vault| is no longer empty.
The shares calculated is $(10 * 1) / 11 = 0$ because of the rounding, which indicates the user does not get any share from his transaction but still transfers tokens at line \ref{line:transfer} in Listing \ref{lst:router}.
After that, the attacker can safely withdraw all 11 tokens because there are no other shareholders.

Generally, \tool can generate the transaction sequence in a few seconds because the intended sequence is short enough, and the second transaction making the donation is exactly what Action A1 defined in Table \ref{tab:actions} does.
Furthermore, the profit function in this vulnerability is monotonous, which is suitable for \tool to pick the most optimal donation amount to drain all victim's assets.
\section{Discussion}


\noindent
\textbf{Improving On-Chain Defense} \tab
Generating attacks and maximizing profits is crucial for on-chain arm-race. For instance, on December 11, 2021, Sortbet Finance was exploited by whitehats to secure 26 million dollars~\cite{sorbet}. As suggested in Section \ref{sec:eval}, \tool has a rather high throughput and supports searching for profits from any given state, either before or after an existing transaction. This enables \tool to potentially defend on-chain smart contracts in two ways: \ding{172} \tool can keep scanning the vulnerable smart contracts locally and notify corresponding projects if there are profitable vulnerabilities. \ding{173} Meanwhile, \tool is capable of watching the pending transactions and front-running the vulnerable transactions to secure users' funds as suggested in Section \ref{sec:rq3}. We think it is another promising application of \tool.

\noindent
\textbf{Automatic Action Synthesis} \tab
While \tool{} has demonstrated its effectiveness and practicality, there are still limitations in terms of its current support for action extraction.
These limitations may affect the effectiveness of testing a broader range of smart contract targets.
Currently, \tool{} supports actions for common liquidation and exchange DeFi applications and concentrates on profitable vulnerabilities. Although these presumptions suit a variety of smart contracts, given the popularity of the actions, these may not apply when testing other targets with different DeFi business models.
Currently, our solution enables users to define custom actions assisted by LLM in Solidity (see more details at our website~\cite{ourwebsite}), obviating the need to modify the fuzzer for enhanced usability.
An interesting area of future research could be to devise adaptive methods that automate this manual process.
Specifically, a domain knowledge model could first learn from the project and then be used to improve the abilities of action and abnormal action.
Previous work, such as ~\cite{actlifter} and ~\cite{defiranger}, has proposed automated ways to extract high-level DeFi semantics from transactions, which can be applied to \tool for better generality.
We consider this an interesting direction for future research.

\noindent
\textbf{Integrate Static Application Security Testing (SAST) Tools} SAST tools like slither~\cite{slither}, mythril~\cite{mythril} and securify~\cite{securify} have been proven effective for auditing solidity projects in both industry and academia. 
SAST tools usually do not require complete deployment for a concrete execution environment, and thus are flexible and scalable to capture a wide range of smart contract vulnerabilities.
However, \tool offers unique advantages compared to SAST tools: \ding{172} Fuzzers can provide concrete reproduction of the vulnerabilities and \tool further ensures the vulnerabilities lead to asset loss without false positives. SAST tools are usually driven by general rules and thus prone to produce false positives. \ding{173} \tool offers unique gradient based exploitation strategy, while SAST tools are usually designed to reveal but not exploit the vulnerabilities.
In addition, SAST tools are complementary rather than exclusive to \tool. For instance, SAST tools can help identify potential fuzzing targets and understand the inter-action dependencies. It is also possible to enhance the proof-of-profit candidate criteria for \tool by SAST tools. We leave the strategic combination of the SAST tools with \tool as future directions.




\section{Related Work}

\noindent
\textbf{Smart Contract Fuzzing} \tab Smart contract fuzzing has been proven an effective way to search for vulnerabilities on Web3. So far, there are many fuzzers adopting strategies borrowed from traditional code coverage-based fuzzing. sFuzz~\cite{sfuzz} implements fuzzing with strategies adapted from AFL for higher code coverage. Smartian~\cite{smartian} proposes both static and dynamic dataflow analysis to reach the vulnerabilities guarded by complex data dependencies. ConFuzzius~\cite{confuzzius} hybrids symbolic execution, data dependency analysis, and taint analysis to exercise deeper bugs. Ethploit~\cite{ethploit} explores generating PoCs using a fuzzing approach. \ityfuzz~\cite{ityfuzz} borrows the concept of dataflow waypoints~\cite{waypoint} to save snapshots to significantly speedup the fuzzing process. In addition, there exist novel attempts to apply learning-based methods in smart contract fuzzing like ILF~\cite{he2019learning}, xFuzz~\cite{xue2022xfuzz}, RLF~\cite{rlf}.

\noindent
\textbf{Profit-Driven Security Researches} \tab
Profitable vulnerabilities on the blockchain are closely linked to factors such as the gap between low-level transactions and high-level DeFi actions, highlighted by tools like DeFiRanger~\cite{defiranger} and ActLifter~\cite{actlifter}, and the detection of anomalous token transfers through fund flow tracking in DeFiWarder~\cite{defiwarder}.
These factors have proven effective in attack detection and practical applications, inspiring \tool to explore their application in smart contract fuzzing.
The concept of Blockchain Extractable Value (BEV), focusing on maximizing profits via strategies like front-running and back-running~\cite{darkforest}, has also been extensively studied.
Works such as APE~\cite{blockimitation} leverage dynamic taint analysis and program synthesis to imitate and front-run victim transactions, while others explore flash loans~\cite{flashloan} or algorithms like Bellman-Ford-Moore~\cite{justintimedisocveryprofit} to identify profitable opportunities.
Although these methods primarily target profit-making strategies without directly exploiting vulnerabilities, they have significantly influenced \tool's development.
Furthermore, given the prevalence of suboptimal DeFi interactions~\cite{subpotimal}, optimizing profits is a meaningful direction.
While \tool employs an SGD-based approach for profit maximization, alternative methods include SMT-based optimizations~\cite{justintimedisocveryprofit} and constraint-solving models tailored to decentralized exchanges and flash loans~\cite{flashloan}.
However, these techniques often depend on smart contracts adhering strictly to specifications, limiting their applicability to mainstream tokens and DEXs in BEV scenarios~\cite{darkforest}.
Challenges still exist for tokens that impose transfer fees or faulty DEXs like \textit{uranium} in our dataset, breaking these constraints and leading to false optima.
Thus, developing adaptive optimization methods for profit maximization remains a valuable avenue for future research.

\noindent
\textbf{Advanced Fuzzing Techniques} \tab
Fuzzing is one of the most practical techniques for detecting zero-day software vulnerabilities~\cite{fuzzing1990}.
There are various efforts dedicated on improving general fuzzing techniques such as grammar-aware fuzzing, hard branch penetration-assisted fuzzing, state-aware fuzzing, fuzz driver generation, fuzzing with domain-specific oracles, etc~\cite{aflfast,Angora,chen2020muzz,li2019cerebro,savior,mopt,ecofuzz,choi2021smartian,aafer2021logfuzz}.
These techniques cannot be directly applied to smart contract fuzzing due to the vast difference between C/C++ projects and smart contracts.
However, they offer valuable insights on improving smart contract fuzzing.

\section{Conclusion}

In this work, we introduced \tool{}, the first profit-centric fuzzing platform designed to detect and maximize the exploitation of profitable vulnerabilities in smart contracts.
By employing innovative techniques such as DeFi action-based mutators, profitability recognition criteria, and gradient descent-based profit maximization, \tool{} effectively addresses the limitations of existing fuzzers.
Our extensive evaluation demonstrated \tool{}'s superior performance over state-of-the-art fuzzers, achieving higher detection rates and maximizing profits, even outperforming real-world highly optimized exploits in some cases.
Furthermore, \tool{} proved valuable to auditors in practical applications by uncovering six zero-day vulnerabilities and earning \$2,500+ bounty rewards.

\section*{Internal Threats to Validity}

The dataset, baselines, and experimental setup in our work could introduce bias, affecting the soundness of our results. Although the dataset was constructed using all publicly known attacks on Uniswap and ERC20 tokens exploited on ETH or BSC over the past two years from DeFiHackLabs~\cite{defihacklabs}, biases could arise from missing or underreported incidents and the overrepresentation of highly publicized attack types. For baselines, we reviewed all known fuzzers from the literature and either included comparison results or explained why comparisons were not feasible. In our experimental setup, we adhered to best practices from prior work~\cite{evalfuzz} to reduce randomness in fuzzing evaluations.

\section*{Data Availability}

To facilitate future research and follow open science policy, we will release both experiment data and supplemental materials of \tool{} on ~\cite{ourwebsite} and ~\cite{zenodo} for archive purpose.

\section*{Acknowlgement}

We gratefully acknowledge the support of this research by the National Research Foundation, Singapore, and the Cyber Security Agency under the National Cybersecurity R\&D Programme (NCRP25-P04-TAICeN).
We also thank the National Research Foundation, Singapore, and DSO National Laboratories under the AI Singapore Programme (AISG Award No: AISG2-GC-2023-008), as well as the National Research Foundation, Prime Minister’s Office, Singapore under the Campus for Research Excellence and Technological Enterprise (CREATE) programme.
Additionally, we acknowledge the support from the National Key Research and Development Program of China (2022YFB2703503).
Any opinions, findings, and conclusions or recommendations expressed in this material are those of the author(s) and do not reflect the views of funding agencies.

\pagebreak

\bibliographystyle{ACM-Reference-Format}
\bibliography{acmart}

\appendix









\end{document}